\documentclass{article}

\usepackage[english]{babel}

\usepackage[letterpaper,top=2cm,bottom=2cm,left=3cm,right=3cm,marginparwidth=1.75cm]{geometry}

\usepackage{amsmath,booktabs}
\usepackage{graphicx}
\usepackage{bbm}
\usepackage[colorlinks=true, allcolors=blue]{hyperref}
\usepackage{float}
\usepackage{subfigure}
\usepackage{lineno,hyperref}
\usepackage{amssymb} 
\usepackage[ruled,vlined]{algorithm2e}
\usepackage{mwe}
\usepackage{caption}
\usepackage{tabularx}
\usepackage[numbers]{natbib}
\usepackage{authblk}

\modulolinenumbers[5]

\title{Advanced Statistical Arbitrage with Reinforcement Learning}
\author[$\dag$,\thanks{Corresponding author. Email: ningb@purdue.edu}]{Boming Ning}
\author[$\dag$]{Kiseop Lee}

\affil[$\dag$]{Department of Statistics, Purdue University}

\providecommand{\keywords}[1]{\textbf{\textit{Keywords---}} #1}

\begin{document}
\maketitle

\begin{abstract}
Statistical arbitrage is a prevalent trading strategy which takes advantage of mean reverse property of spread of paired stocks. Studies on this strategy often rely heavily on model assumption. In this study, we introduce an innovative model-free and reinforcement learning based framework for statistical arbitrage. For the construction of mean reversion spreads, we establish an empirical reversion time metric and optimize asset coefficients by minimizing this empirical mean reversion time. In the trading phase, we employ a reinforcement learning framework to identify the optimal mean reversion strategy. Diverging from traditional mean reversion strategies that primarily focus on price deviations from a long-term mean, our methodology creatively constructs the state space to encapsulate the recent trends in price movements. Additionally, the reward function is carefully tailored to reflect the unique characteristics of mean reversion trading.

\end{abstract}



\section{Introduction}
\label{sec1}
Statistical arbitrage, also known as mean reversion trading or pairs trading, is an important trading strategy in the financial markets. The essence of statistical arbitrage lies in creating spreads or portfolios from the market that exhibit mean-reverting characteristics, thereby unlocking opportunities for profit. For instance, if the price of a spread falls below its long-term mean, a trader might take a long position and then wait until its price correction, aiming to profit from this adjustment.

The approach to statistical arbitrage unfolds in three distinct steps: First, it entails the identification of two or more securities that have shown a historical pattern of moving together. Next, a mean-reverting spread is formulated from these correlated securities. The final step involves taking a position when the spread deviates from its long-term mean, leveraging the anticipated return to equilibrium to generate profits. Therefore, mean reversion trading is divided into three main elements: (1) the identification of securities with co-movements, (2) the construction of mean-reverting spreads, and (3) the development of a trading strategy based on these mean-reverting spreads. The first two components are referred to as the formation phase, while the third is considered the trading phase.

The initial step in statistical arbitrage strategy is the identification of similar securities.  Traditional methods predominantly utilize distance metrics, as highlighted by \citeauthor{gatev2006pairs} (\citeyear{gatev2006pairs}), where pairs are formed by selecting the securities that minimizes the sum of squared deviations (SSD) between their normalized price series. This principle of pair selection can be extended to encompass pairs of representative stocks and ETFs within specific sectors (\citeauthor{gatev2006pairs} (\citeyear{gatev2006pairs}), \citeauthor{avellaneda2010statistical} (\citeyear{avellaneda2010statistical}), \citeauthor{Montana2011} (\citeyear{Montana2011}), \citeauthor{meanReversionBook} (\citeyear{meanReversionBook})), as well as physical commodities and their corresponding stocks/ETFs (\citeauthor{kanamura2010profit} (\citeyear{kanamura2010profit})) and entities within the cryptocurrency market (\citeauthor{leung2019constructing} (\citeyear{leung2019constructing})). These references highlight the adaptability and efficacy of statistical arbitrage strategies across a broad spectrum of markets. In our study, we select ten representative pairs from various sectors within the US market to construct the mean reversion portfolios.

After identifying groups of similar stocks, the next step involves constructing an statistical arbitrage portfolio or spread with mean-reverting property. A foundational approach, as suggested by \citeauthor{gatev2006pairs} (\citeyear{gatev2006pairs}), involves taking a long position in one security of the pair and a short position in the other, that is, trading on the spread \(S_1 - S_2\) for two similar stocks \(S_1\) and \(S_2\). Although straightforward, this method often cannot create an optimal portfolio that exhibits high mean-reverting characteristics. A more sophisticated strategy frequently used is Ornstein–Uhlenbeck (OU) mean reversion trading (\citeauthor{meanReversionBook} (\citeyear{meanReversionBook})). For a pair of similar stocks \(S_1\) and \(S_2\), the goal is to find a coefficient \(B\) such that the spread \(S_1- B \cdot S_2\) mimics an OU process as closely as possible, with \(B\) typically determined through maximum likelihood estimation based on the OU process distribution. However, the real-world application of this strategy faces challenges due to the fact that financial markets may not always align with the assumptions of the Ornstein–Uhlenbeck process, which can compromise the effectiveness of strategies based on this model.

To overcome the limitations inherent in these assumption-dependent methods, we introduce a novel approach that utilizes the proposed empirical mean reversion time of any time series as a measure of reversion speed. This allows for the construction of a mean reversion spread without relying on any theoretical assumptions. By employing a grid search method, we can systematically explore different combinations to identify an optimal spread that possesses the least empirical mean reversion time. This technique offers a more flexible and potentially more robust framework for arbitrage portfolio construction.

The final phase of statistical arbitrage involves formulating a trading strategy based on the constructed mean-reverting spread. Traditional strategies heavily rely on model parameter estimations. For instance, \citeauthor{gatev2006pairs}(\citeyear{gatev2006pairs}) initiate a trade when a spread's price deviation exceeds two historical standard deviations from the mean, calculated during the pair formation phase, and exit the trade upon the next convergence of prices to the historical mean. In the context of Ornstein-Uhlenbeck (OU) mean reversion trading, parameter estimations for the long-term mean and volatility of the OU model are typically employed to define trading criteria (\citeauthor{leung2015optimal} (\citeyear{leung2015optimal}), \citeauthor{meanReversionBook} (\citeyear{meanReversionBook})). These estimations depend on historical data from the formation period, with an underlying assumption that parameters remain constant in the following trading phase—an assumption that may not hold due to market fluctuations. Additionally, the selection of hyper-parameters significantly impacts trading performance, yet a robust method for the optimal hyper-parameters selection remains absent. For example, the determination of an appropriate threshold for price deviation from the mean lacks a clear consensus. 

To address these challenges, we introduce a reinforcement learning (RL) algorithm designed to dynamically optimize trading decisions over time, replacing the need for predefined rules. This approach models the task within a reinforcement learning framework, aimed at enabling agents to take actions that maximize cumulative rewards in an environment. We design the state space to capture the recent movements of the spread price, thus moving away from a dependence on historical mean and standard deviation estimates. This approach enables the agent to make informed decisions about future actions by leveraging insights into current market trends, rather than depending on parameter estimations from the formation period. We get the rid of the hyper-parameters choice at the same time.
Simultaneously, we remove the necessity for hyper-parameter selection by not incorporating universal hyper-parameters, such as thresholds, which can significantly impact trading performance. This approach streamlines the trading process, focusing on dynamic adaptation without the constraints of fixed parameters.

The structure of the paper is organized as follows. Section 2 provides an overview of related research in the field. Section 3 details the definition of empirical reversion time for spreads and outlines the methodology for identifying optimal asset coefficients by minimizing mean reversion time. Section 4 presents a reinforcement learning framework designed for the development of optimal trading strategies. Experimental results, based on simulated data and real-world applications in the US stock market, are discussed in Section 5. Finally, Section 6 offers conclusions and outlines future research directions.

\section{Related Research} 
\label{sec2}
The seminal work by \citeauthor{gatev2006pairs} (\citeyear{gatev2006pairs}) marks a cornerstone in the study of pairs trading, a strategy predicated on the mean reversion principle. By employing what is now known as the Distance Method (DM), they analyzed CRSP stocks from 1962 to 2002, identifying trading opportunities when the price of asset pairs deviated beyond two historical standard deviations and closing positions upon price convergence. This approach yielded an excess return of 1.3 $\%$ for the top 5 pairs and 1.4$\%$ for the top 20 pairs. Building on this, \citeauthor{do2012pairs} (\citeyear{do2012pairs}) further examined the viability of pairs trading considering transaction costs. Their findings enhance the understanding of pairs trading's practical application, demonstrating its feasibility even when accounting for trading expenses.

Beyond the Distance Method, the stochastic spread method emerges as a significant alternative for mean reversion trading, utilizing stochastic processes to analyze the mean-reverting nature of spreads. This approach involves constructing spreads and generating trading signals based on the analysis of parameters within the chosen stochastic model. \citeauthor{elliott2005pairs} (\citeyear{elliott2005pairs}) were pioneers in this area, introducing a Gaussian Markov chain model to capture the mean-reverting dynamics of spreads. They leveraged model estimates against observed spread data for trading decisions, laying the groundwork for further exploration. Building on this, \citeauthor{do2006new} (\citeyear{do2006new}) expanded the concept with a generalized stochastic residual spread method, aimed at modeling relative mispricing more comprehensively. Their approach broadened the stochastic spread methodology's applicability in mean reversion trading. Further enriching this field, the extensive work by \citeauthor{meanReversionBook} (\citeyear{meanReversionBook}) delves into optimal mean reversion trading strategies based on various stochastic models. Covering models such as the Ornstein-Uhlenbeck, Exponential OU, and CIR, this research illuminates the versatility and effectiveness of stochastic models in optimizing trading strategies. This progression of work significantly advances our understanding of mean reversion trading by demonstrating the potential of diverse stochastic approaches.

Cointegration tests stand as a critical alternative method for mean reversion trading strategies. Leveraging the foundational error correction model introduced by \citeauthor{engle1987co} (\citeyear{engle1987co}), \citeauthor{vidyamurthy2004pairs} (\citeyear{vidyamurthy2004pairs}) outlines a cointegration framework that has become essential in pairs trading analysis. This methodology is advanced by \citeauthor{galenko2012trading} (\citeyear{galenko2012trading}), who develops active trading strategies for ETFs, utilizing cointegration to exploit trading opportunities within exchange-traded funds. Furthermore, \citeauthor{huck2015pairs} (\citeyear{huck2015pairs}) conducts a comparative analysis, examining the performance of the Distance Method against cointegration-based strategies within the S$\&$P 500, clarifying the relative merits of these methodologies in the context of mean reversion trading. Demonstrating the method's extensive applicability, \citeauthor{leung2019constructing} (\citeyear{leung2019constructing}) crafts cointegrated cryptocurrency portfolios using both the Engle-Granger two-step approach and the Johansen cointegration test, highlighting cointegration's adaptability across different asset classes.  

In recent years, the landscape of mean reversion trading has been enriched by a variety of innovative methods. Among these, the use of copulas has gained attention for its ability to model the dependence between asset pairs, as evidenced by the work of \citeauthor{liew2013pairs} (\citeyear{liew2013pairs}) and \citeauthor{xie2016pairs} (\citeyear{xie2016pairs}). Additionally, an optimization approach has been explored by \citeauthor{zhang2020sparse} (\citeyear{zhang2020sparse}), who seek to construct sparse portfolios with mean-reverting price behaviors from multiple assets. Machine learning techniques have also emerged as a powerful tool in this domain. With contributions from \citeauthor{guijarro2021deep} (\citeyear{guijarro2021deep}), \citeauthor{sarmento2020enhancing} (\citeyear{sarmento2020enhancing}), and \citeauthor{chang2021pairs} (\citeyear{chang2021pairs}), machine learning algorithms have been demonstrated to be able to uncover complex statistical patterns and relationships among assets. These developments signal a period of significant innovation in statistical arbitrage, providing traders and researchers with an expanded toolkit for strategy development and implementation.

\section{Empirical Mean Reversion Time: Spread Construction}
\label{sec3}
Once we identify groups of similar stocks, we need to form an arbitrage portfolio with mean reversion property. In the traditional OU pairs trading, for two similar stocks $S_1$ and $S_2$, we choose $B$ such that the spread $S_1- B S_2$ follows an OU process as closely as possible. A reasonable way to find $B$ is to use the maximum likelihood estimator, using the distribution of the OU process. However, since we do not have a model assumption, we cannot use MLE anymore.

We extend paired trading to a multi-asset portfolio. In other words, given $n$ similar stocks $S_i, i=1,2,...,n$, we form a spread
$X = \sum_{i=1}^n a_i S_i.$
Our goal is to find a portfolio $(a_1, a_2,...,a_n)$ such that the spread $Y$ has a mean reverting property as much as possible.

Let us consider a popular OU process
\[
dX_t = \mu (\theta - X_t) dt + \sigma dW_t,
\]
where $W_t$ is a standard Brownian motion. An empirical result shows that $\mu$ has the biggest impact on the profit among three parameters $\mu, \theta$ and $\sigma$. In general, a larger $\mu$ gives higher return. Intuitively, it should be the case, since a larger $\mu$ implies a faster mean reversion. Therefore, a trader can make a quick profit by taking advantage of the deviation from the mean.

Therefore, we want to make the spread $X=\sum_{i=1}^n a_i S_i$ to have a faster mean reversion property. Consider a trading strategy to buy at $X_t=\theta-a$ and sell later at $X_t=\theta$ for $a>0$ and a given long-term mean $\theta$. 
Define the stopping time
\begin{equation}\label{equ3}
\tau_t = \inf\{s>t: X_s=\theta\ |\ X_t = \theta - a \}.
\end{equation}
A faster mean reversion corresponds to a smaller $\tau_t$.  
Based on this logic, it is natural to define our arbitrage portfolio selection problem as follows.

\emph{At time $t$, consider the training time interval $[t-h,t]$. Find the optimal portfolio $(a_1, a_2,...,a_n)$ which minimizes the sample mean of $\tau$'s in this interval, given $\bar{Y}=\theta$ and $S^2 (Y) < M$ for a constant M.}

The reason we impose the upper bound $M$ on the sample variance is because of the constraint of the initial wealth and to prevent a large leverage, which makes the portfolio unstable.

\subsection{Empirical Mean Reversion Time}
In this part, we introduce the conception of empirical mean reversion time based on the idea above.

Inspired by \citeauthor{fink2007important} (\citeyear{fink2007important}), we firstly define the important extremes of time series. Let $s$ be the sample standard deviation of a time series $X_t$, where $t\in[0, T]$. In real market, we can only get the discrete data points. So we assume a time series $(X_1,\cdots,X_n)$. Let $C$ be a positive constant. A point $X_m$ is an important minimum of the time series if there are indices $i$ and $j$, where $i \leq m \leq j$, such that
\begin{itemize}
    \item $X_m$ is the minimum among $X_i,\cdots,X_j$;
    \item $X_i-X_m \geq C\cdot s$ and $X_j-X_m \geq C\cdot s$.
\end{itemize}
Intuitively, $X_m$ is the minimal value of some segment $X_i,\cdots,X_j$, and the endpoint values of this segment are much larger than $X_m$. Similarly, $X_m$ is an important maximum if there are indices $i$ and $j$, where $i\leq m \leq j$, such that
\begin{itemize}
    \item $X_m$ is the maximum among $X_i,\cdots,X_j$;
    \item $X_m - X_i \geq C\cdot s$ and $X_m-X_j \geq C\cdot s$.
\end{itemize}

Drawing on the conceptual framework introduced by Equation (\ref{equ3}), our objective is to propose an empirical mean reversion time that quantifies the duration required for the spread to revert to its long-term mean, starting from the maximum deviation observed. It enables us to infer the optimal coefficients for securities by minimizing the spread's empirical reversion time.

We now proceed to construct a sequence of time moments \(\{\tau_i\}_{i=0}^{N}\), derived recursively from the significant local extremes within the actual asset price process.
More precisely, we define the initial time moment as
\[
\tau_1 = \inf\{u \in [0, T] : X_u \text{ is a local extreme}\}.
\]
Subsequently, $\tau_2$ is identified as the first instance when the series crosses the sample mean $\hat{\theta}$, defined by
\[
\tau_2 = \inf\{u \in [\tau_1, T] : X_u = \hat{\theta}\}.
\]
Recursively, $\tau_3$ is the first local extreme following $\tau_2$, and $\tau_4$ is the first crossing of the long-term mean after $\tau_3$, and so on. Thus, all odd-numbered time moments $\{\tau_n\}_{n=1,3,5,\cdots}$ correspond to local extremes and are defined as
\[
\tau_n = \inf\{u \in [\tau_{n-1}, T] : X_u \text{ is a local maximum}\}.
\]
Conversely, all even-numbered time moments $\{\tau_n\}_{n=2,4,6,\cdots}$ are associated with the crossings of the long-term mean, specified by
\[
\tau_n = \inf\{u \in [\tau_{n-1}, T] : X_u = \hat{\theta}\}.
\]
The complete sequence $\{\tau_n\}_{n=1}^{N}$ is constructed in an inductive manner.

Once we get the time stamps of iterated time stamps $\{\tau_n\}$, the empirical reversion time $r$ is defined as the average of the time interval from local extremes to crossing times. That is,
$$
r = \frac{2}{N} \sum_{\substack{i=2 \\ i \, \text{even}}}^{N} (\tau_n - \tau_{n-1})  
$$

Next, we briefly introduce a \textit{grid search algorithm} that can help us find the optimal coefficients based on the empirical mean reversion time.
Assume the price processes of $n$ similar assets are denoted by $S_1, S_2, \ldots, S_n$. Our aim is to find the optimal coefficients $(a_1, a_2, \ldots, a_n)$ such that the portfolio $X=\sum_{i=1}^n a_i S_i$ exhibits the minimal empirical mean reversion time. Without loss of generality, we set the first coefficient to $a_1 = 1$. We then evaluate the empirical mean-reversion time of $Y$ for each coefficient $a_i$, where $a_i \in [-3.00, -2.99, -0.98, \ldots, 2.99, 3.00]$ for $2 \leq i \leq N$. The optimal coefficients are determined by selecting the set that minimizes the empirical mean reversion time of $Y$.

\section{Reinforcement Learning: Advanced Trading Strategies}
\label{sec4}
The final stage of statistical arbitrage involves developing a trading strategy based on a mean-reverting spread. Traditional approaches assume parameters' stability from formation phase to trading phase, which market changes can challenge. Moreover, the choice of hyper-parameters, such as the deviation threshold, critically affects performance, yet a standard method for their optimal selection is lacking.

Our motivation is to leverage reinforcement learning algorithms to help us decide the optimal trading actions dynamically over time, other than design some preset rules manually. Reinforcement learning framework is a machine learning method concerned with how intelligent agents ought to take optimal actions in an environment in order to maximize the cumulative reward. 

\subsection{Preliminaries of Reinforcement Learning}
In RL, the sequential decision-making problem is modeled as Markov decision process (MDP), which is an augmented structure of Markov process. In addition to a Markov process, one has the possibility of choosing an action from an available action space and get some reward that tells us how good our choices were at each step. 

The “environment” is defined as the part of the system outside of
the RL agent’s control. At each time step $t$, we observe the current
state of the environment $S_t \in \mathcal{S}$ and then chooses an action $A_t \in \mathcal{A}$. The choice of action influences both the transition to the next state, as well as the reward received, $R_t$. Thus, we will get a sequence in MDP as:
$$
S_0, A_0, R_1, S_1, A_1, R_2, S_2, A_2, R_3, \cdots
$$
Every MDP is uniquely determined by a multivariate conditional probability distribution $p(s', r|s, a)$, which is the joint probability of transitioning to state $s'$ and receiving reward $r$, conditional on the previous state being $s$ and taking action $a$.

A policy $\pi$ is a mapping from states to probability distributions over the action space. If the RL agent is following policy $\pi$, then in state $s$ it will choose action a with probability $\pi(a|s)$. To find the optimal policy, one must specify a goal function. A wide-used discounted goal function is defined as
\begin{equation}
    \begin{aligned}
        G_t &= R_{t+1} + \gamma R_{t+2} + \gamma^2 R_{t+3} + \cdots \\
            &= R_{t+1} + \gamma G_{t+1},
    \end{aligned}
\end{equation}
where $R_t$ is the instant reward at time $t$ and $\gamma \in (0, 1)$ is a discount factor expressing that rewards further in the future are worth less than rewards which are closer in time. Our goal is to search for the optimal policy that maximizes the expectation of the goal function, namely
$$
\max_\pi \mathbb{E}[G_t]
$$

Next we introduce related concepts of Q-learning. The action-value function for policy $\pi$ is the expectation of goal function, assuming we start in state $s$, take action $a$ and then follow the policy $\pi$ from then on
$$
q_\pi(s, a) := E[G_t|S_t = s, A_t = a].
$$
The optimal action-value function is then defined as
$$
q_*(s,a) = \max_\pi q_\pi(s, a).
$$
If we knew the optimal action-value function, we would know the optimal policy itself easily, that is, choose $a \in \mathcal{A}$ to maximize $q_*(s, a)$. 
Hence we can reduce the problem to finding $q_*$, which is solved iteratively based on the Bellman equations. It is straightforward to establish the Bellman equation of action-value function:
\begin{equation}
    q_*(s,a) =  \sum_{s', r} p(s',r|s,a)[r+\gamma \max_{a'} q_*(s',a')].
\end{equation}
The core of the Q-learning is to leverage Bellman equation as a simple value iteration update, using the weighted average of the old value and the new information.

Before learning begins, the approximate action-value function Q is initialized to a possibly arbitrary value. Then, the corresponding sample update for q-function of $S_t, A_t$, given a sample next state and instant reward, $S_{t+1}$ and $R_{t+1}$ (from the model), is the Q-learning update:
\begin{equation}\label{equ6}
    Q^{new}(S_t, A_t) \leftarrow Q(S_t, A_t) +\alpha \cdot \left(R_{t+1} + \gamma \cdot \max_a Q(S_{t+1}, a) - Q(S_t, A_t) \right),
\end{equation}
where $\alpha$ is the learning rate, $\gamma$ is the discount factor, $R_{t+1}$ is the reward received after taking action $A_t$ in state $S_t$, and $S_{t+1}$ is the next state. Note that $Q^{new}(S_t, A_t)$ is the sum of three factors:
\begin{enumerate}
    \item $(1-\alpha)Q(S_t,A_t)$: the current value weighted by the learning rate.
    \item $\alpha \cdot R_{t+1}$: the weighted instant reward to obtain if action $A_t$ is taken when in state $S_t$.
    \item $\alpha \cdot \gamma \cdot \max_a Q(S_{t+1}, a)$: the maximum cumulative reward that can be obtained from next state $S_{t+1}$ (weighted by learning rate and discount factor)
\end{enumerate}
Generally, $R_{t+1} + \gamma \cdot \max_a Q(S_{t+1}, a)$ is referred as target $Y_t$. Thus, the iteration (\ref{equ6}) updates the current value $Q(S_t, A_t)$ towards a target value $Y_t$. 

An epsilon-greedy strategy is employed for action selection. In the training phase, actions are chosen at random with a probability of $\epsilon$, whereas the action with the highest Q-value is selected with a probability of $1 - \epsilon$. This approach facilitates a balance between exploration of new actions and exploitation of known values. In the testing phase, when the trained agent is assessed using new incoming data, $\epsilon$ is adjusted to $0$. This modification ensures that action selection is solely based on the highest Q-value, thereby focusing entirely on exploitation based on the acquired knowledge. Thus, the model incorporates both exploration and exploitation during training, while adopting a strategy of pure exploitation during testing.

\subsection{RL Model for Mean Reversion Trading}
Now we can introduce our reinforcement learning model for an optimal mean reversion trading strategy.

The state space is constructed based on the trajectory of price movements over a recent sequence of time points. At any particular moment \(t\), the state, denoted \(S_t\), is encapsulated by the vector
\[
S_t = [d_{t-l+1}, d_{t-l+2}, \ldots, d_{t}],
\]
where each \(d_{i}\) characterizes the direction and magnitude of price changes at time \(i\). A positive value of \(d_{i}\) signifies a price increase relative to time \(i-1\), and conversely, a negative value indicates a decline. The magnitude of \(d_{i}\) quantifies the extent of this change. Formally, for each \(i\) within the interval \(t-l+1 \leq i \leq t\), let \(\pi_i = \left( \frac{P_i - P_{i-1}}{P_{i-1}} \right) \times 100\) represent the percentage price change from \(i-1\) to \(i\). The definition of states is given by
\[
  d_i = 
  \begin{cases} 
    +2 & \text{if } \pi_i > k, \\
    +1 & \text{if } 0 < \pi_i < k, \\
    -1 & \text{if } -k < \pi_i < 0, \\
    -2 & \text{if } \pi_i < -k.
  \end{cases}
\]
Here, '+2' indicates a significant increase, '+1' a moderate increase, '-2' a significant decrease, and '-1' a moderate decrease. The choice of threshold \(k\), such as 3\%, is adjustable to accommodate different sensitivity levels. This approach results in a state space comprising \(4^l\) unique states, effectively capturing a wide spectrum of recent price movement scenarios.

The action space is composed of three possible actions: selling one share is represented by $-1$, taking no action is denoted by $0$, and buying one share is indicated by $+1$. The set of available actions at any given time is contingent upon the agent's current position. Specifically, when the agent does not hold any position, the permissible actions include buying ($+1$) or holding ($0$). Conversely, if the agent is currently in a long position, the options are limited to selling ($-1$) or holding ($0$). Note that our model does not account for initially entering a short position.

The immediate reward, $R_{t+1}$, earned by the agent for taking action $A_t$ under prevailing environmental conditions, is mathematically defined as:
\begin{equation}\label{equ7}
    R_{t+1} = A_t \cdot (\theta - X_t) - c \cdot |A_t|,
\end{equation}
where $X_t$ denotes the current price of the spread, and $\theta$ represents the true global mean of $X_t$. The formulation is designed such that a buy action ($A_t = +1$) is rewarded positively when $X_t$ is below its long-term mean, $\theta$, encouraging purchases at lower prices. Conversely, a sell action ($A_t = -1$) incurs a negative reward under the same conditions. If $X_t$ exceeds the long-term mean, resulting in a negative value for $\theta - X_t$, the rewards for buy and sell actions are adjusted accordingly to discourage buying at high prices and encourage selling. The term $c$ represents the transaction cost per trade.

The cumulative return from time $t$ to the terminal time $T$ is expressed as:
\begin{equation}
    G_t = \sum_{s=t+1}^{T} e^{-r \cdot (s-t)} \cdot R_s + I_T \cdot X_T,
\end{equation}
where $r$ denotes the interest rate, reflecting the time value of money, and $I_T$ signifies the position held at the terminal time. This formulation accounts for the exponential decay of rewards over time due to the discounting effect of the interest rate, emphasizing the importance of immediate gains and the impact of holding a position until the end of the considered period.

To accurately fit the optimal Q-table, ample training data is essential. However, the real market offers only a limited observation path of spreads, presenting a significant challenge for effective training. Additionally, our reward function, as defined in Equation (\ref{equ7}), incorporates the true long-term mean of the spread—a value that remains elusive in actual market scenarios. 

To overcome these obstacles, our strategy involves initially simulating a multitude of mean reversion spreads with different parameters to train the reinforcement learning (RL) agent. This step allows for extensive exposure to various market conditions, enhancing the agent's learning and decision-making capabilities. Subsequently, the model, now adept from the simulation training, is applied to execute trades in the real market. In the language of RL, this method entails utilizing a simulated environment for the agent's training phase.

\section{Experiments}
\label{sec5}
In this chapter, the performance of the proposed method is thoroughly evaluated. Initially, tests are conducted on simulated data to assess the efficacy of the proposed empirical mean reversion time, and the reinforcement learning (RL) model. Subsequently, mean reversion trading experiments are carried out on the S$\&$P 500 using the proposed model-free framework, with outcomes compared against those achieved with other classical statistical arbitrage methods.

\subsection{Empirical Mean Reversion Time}
In this section, we investigate the empirical mean reversion time by conducting simulations of the Ornstein-Uhlenbeck (OU) process. We fix the parameters $\theta = 0$ and $\sigma = 1$, and vary $\mu$ from 2 to 20 in increments of 2. For each parameter combination, we simulate 100 paths of the OU process, each with a terminal time of $T=1.0$ and $n=1000$ data points. Subsequently, we calculate the average empirical mean reversion time using a threshold of $C=2$ for these paths. The primary objective is to compute the average empirical mean reversion time for these paths using a threshold value of \(C=2\), thereby examining the impact of the mean reversion parameter \(\mu\) on the empirical mean reversion time.

\begin{figure}
    \centering
    \includegraphics[width=10 cm]{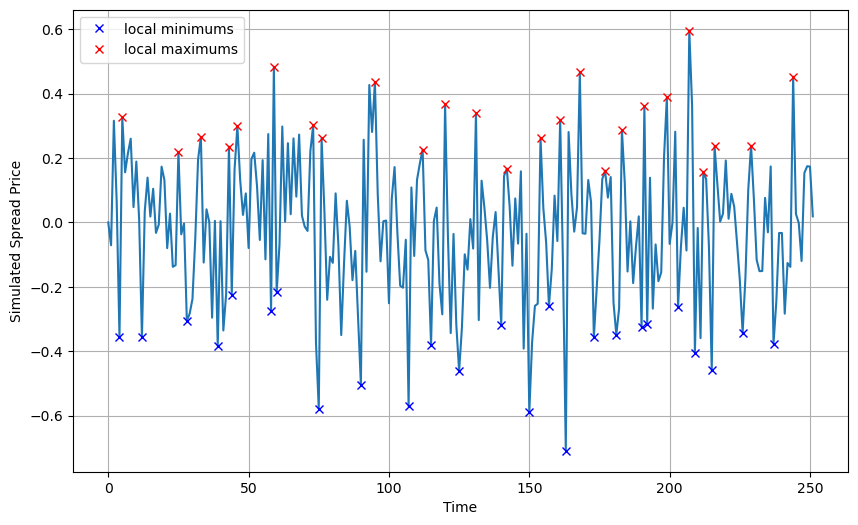}
    \caption{Local extremes calculated on a simulated OU spread with $\theta = 10$, $\theta = 0$ and $\sigma = 1$.}
    \label{fig1}
\end{figure} 

Figure~\ref{fig1} presents the identified local extremes on a simulated Ornstein-Uhlenbeck (OU) path characterized by parameters $\mu = 10$, $\theta = 0$, and $\sigma = 1$. The analysis reveals that the defined criteria for extreme points are highly effective in pinpointing nearly every instance where the time series reaches a local maximum or minimum. This capability underscores the precision of our approach in capturing significant turning points within the simulated path, providing a robust method for analyzing mean reversion characteristics. The accurate identification of these extremes is critical for the development of mean reversion time.

\begin{table}
\centering
\begin{tabular}{cccc}
\toprule
Parameter $\mu$ & Average EMRT & Parameter $\mu$ & Average EMRT \\
\midrule
2.0 & 98.79 & 12.0 & 49.22 \\
4.0 & 83.45 & 14.0 & 45.10 \\
6.0 & 78.09 & 16.0 & 38.04 \\
8.0 & 59.22 & 18.0 & 35.63 \\
10.0 & 58.51 & 20.0 & 31.15 \\
\bottomrule
\end{tabular} 
\caption{Variation of average empirical mean reversion time (EMRT) with parameter $\mu$ in the Ornstein-Uhlenbeck Process.}\label{tab:1} 
\end{table}

The results of the impact of the mean reversion parameter \(\mu\) on the empirical mean reversion time are succinctly presented in Table~\ref{tab:1}, which supports our initial hypothesis by demonstrating a clear inverse relationship between the mean reversion speed, \(\mu\), and the empirical mean reversion time. As \(\mu\) increases, the mean reversion time decreases, indicating a faster adjustment of the process back to its mean level. This finding confirms our hypothesis that the empirical mean reversion time reflects the mean-reverting speed of financial time series.

\subsection{RL Trading on Simulated Data}
In this part, we introduce a preliminary simulated experiment into the effectiveness of the proposed reinforcement learning strategy tailored for mean reversion trading. With the parameters fixed at \(\mu = 1\), \(\theta = 1\), and \(\sigma = 0.1\), we simulate 10,000 paths of the Ornstein-Uhlenbeck process. Each path is designed to reach a terminal time of \(T=252\) and includes \(n=252\) data points, effectively simulating one year of data for use as training samples in our study. This process was chosen due to its relevance in modeling mean-reverting financial instruments, thereby providing a realistic and challenging environment for training our reinforcement learning model. 

Our reinforcement learning (RL) model configuration employs a lookback window size of \(l=4\), generating \(16\) distinct states. Hyper-parameters are set with a learning rate of \(0.1\), a discount factor of \(0.99\), an epsilon of \(0.1\) for the epsilon-greedy strategy and \(10\) training episodes.

\begin{figure}
    \centering
    \includegraphics[width=10 cm]{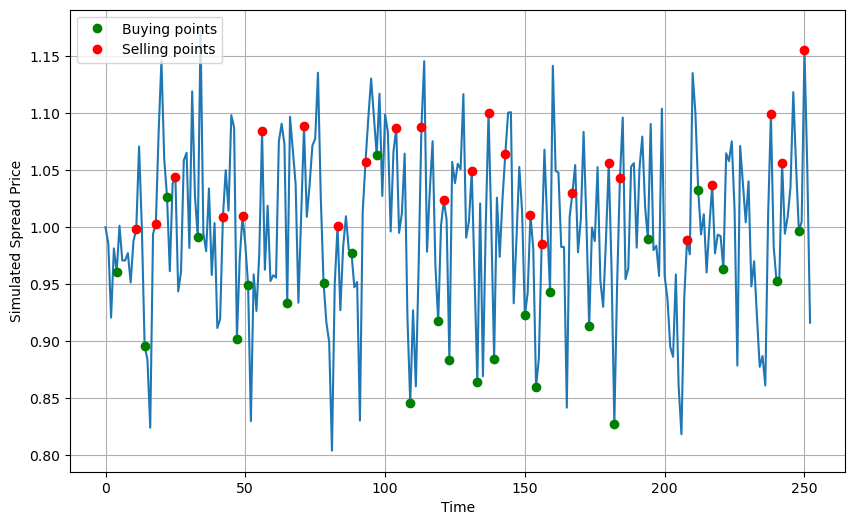}
    \caption{Simulated trading on OU process based on reinforment learning framework.}
    \label{fig2}
\end{figure}

Following the training phase, we apply the trained model to a new, distinct OU sample path to evaluate its decision-making power. This simulated trading are visually presented in Figure~\ref{fig2}, where buy and sell actions executed by the RL agent are denoted by green and red points, respectively. The outcome demonstrates that the trained agent is capable of executing a series of strategic buy and sell decisions.

Subsequently, we evaluate the trained reinforcement learning model on 100 new samples, calculating the average accumulated profits across these samples. An initial investment of 100 dollars is allocated to each new OU sample. At each purchase point recommended by the RL agent, the entire available cash is used to take a long position on the spread. These positions are then closed at the selling points suggested by the agent. The simulation results in an average profit exceeding 600\% across these 100 paths, underscoring the model's adeptness at identifying and leveraging trading opportunities following its training period.

\subsection{Real World Experiments}
In this section, we conduct real-world experiments on S$\&$P500 to evaluate the performance of our proposed strategy, comparing it against established benchmarks such as the classic distance method (DM) \cite{gatev2006pairs} and the Ornstein-Uhlenbeck (OU) mean reversion trading strategy \cite{avellaneda2010statistical, meanReversionBook}. 

\subsubsection{Benchmarks}
We begin by introducing the details of the implementation of these benchmark strategies.

For the distance method \cite{gatev2006pairs}, the initial step involves calculating the sum of squared deviations among all potential pairs' normalized price series. This is followed by the identification and selection of pairs of securities that yield the minimum sum of squared deviations. Subsequently, a mean-reverting spread is constructed, denoted as \(X = S_1 - S_2\), where \(S_1\) and \(S_2\) represent two analogous stocks. In this context, a long position is assumed in one security of the pair, while a short position is taken in the other. Additionally, estimates of the long-term mean and standard deviations are determined during the formation period. 

Transitioning to the trading phase of distance method, the strategy prescribes initiating a long position when the spread's price deviation falls below multiples of estimated standard deviations from the long-term mean. The trade is then exited upon the subsequent reversion of prices. To be more clear, we clarify the trading criterion that we use in our experiment:
\begin{itemize}
    \item buy to open if \(X_t - \Bar{x} < -k\cdot s\)
    \item close long position if \(X_t - \Bar{x} > k\cdot s\)
\end{itemize}
where $\Bar{x}$ and $s$ are the sample mean and standard deviance estimated from the formation period. The threshold parameter \(k\) is set to 1 in our experiment, 
 Note that we solely focuses on the initial engagement in a long position within the portfolio.

For the OU mean reversion trading strategy \cite{avellaneda2010statistical, meanReversionBook}, we also select the
pairs of securities that yield the minimum sum of squared deviations. The next step consists of constructing mean-reverting spreads for the pairs, denoted as \(X = S_1 - B\cdot S_2\), where \(S_1\) and \(S_2\) represent two analogous stocks and \(B\) is determined by maximizing the likelihood score of fitting the spread to an OU process. Furthermore, the parameters of the spreads are estimated as an OU process, including the mean reversion speed \(\hat{\mu}\), the long-term mean \(\hat{\theta}\), and the volatility \(\hat{\sigma}\). 

In the OU trading phase, the equilibrium variance is calculated as \(\hat{\sigma}_{eq} = \frac{\hat{\sigma}}{\sqrt{2\hat{\mu}}}\), according to \citeauthor{avellaneda2010statistical} (\citeyear{avellaneda2010statistical}). The basic trading signals are based on the estimations of the OU parameters: 
\begin{itemize}
    \item buy to open if \(X_t - \hat{\theta} < -k\cdot \hat{\sigma}_{eq}\)
    \item close long position if \(X_t - \hat{\theta} > k\cdot \hat{\sigma}_{eq}\)
\end{itemize}
where \(k\) represents the cutoff value and we set it to \(0.5\) in our experiment. The trading remains exclusively on initially entering a long position in the portfolio.

\subsubsection{Data}
Our experimental framework is designed to encompass a one-year formation period, subsequently followed by a trading period spanning the subsequent year. We utilize the daily adjusted closing prices of representative stocks from different sectors within the U.S. market to construct mean reversion spreads. Our selection includes pairs such as MSFT-GOOGL from Technology, CVS-JNJ from Healthcare, CL-KMB from Consumer Goods, V-MA from Financials, GE-BA from Industrials, OXY-XOM from Energy, WELL-VTR from Real Estate, PPG-SHW from Materials, VZ-TMUS from Telecommunication, and CSX-NSC from Transportation. Data on the daily closing prices for these stocks was collected over the period from January 1, 2022, to December 31, 2023. The data was sourced from the Yahoo! Finance API\footnote{\url{https://pypi.org/project/yfinance/}}.
   
Following the completion of the annual trading, we will collate and analyze the data to calculate the trading performance across various sectors. This process is designed to rigorously evaluate the efficacy of our strategy across different market segments, thereby demonstrating its consistency and adaptability in the face of financial market uncertainties. 

\subsubsection{Experimental Results}

In the forthcoming part, we delve into a detailed analysis of a one-year study, with 2022 designated as the formation period and 2023 as the trading period. During the formation phase, we construct mean reversion portfolios denoted as \(X = S_1 - B\cdot S_2\), where \(S_1\) and \(S_2\) symbolize the first and second stocks, respectively, as listed above. Here, \(B\) represents a positive coefficient tailored to each trading strategy. Specifically, for the Distance Method, this coefficient is uniformly set to 1 across all pairs. In contrast, for OU pairs trading, \(B\) is determined by optimizing the likelihood score of the pair's fit to an Ornstein-Uhlenbeck (OU) process. Within our proposed methodology, \(B\) is calibrated by aiming to minimize the empirical mean reversion time of the spread. Table~\ref{tab2} compiles the pairs trading coefficient \(B\) for each selected pair's mean reversion portfolio, comparing the benchmarks with our novel method. 
\begin{table}
\centering
\begin{tabular}{c|ccc}
\toprule
                   & & Pairs Trading Coefficient $B$& \\
Pairs Index & DM & OU & EMRT\\
\midrule
MSFT-GOOGL    &   1.0 & 0.99  &  0.89 \\
CVS-JNJ       &   1.0 & 0.43  & -0.24 \\
CL-KMB        &   1.0 & 0.39  &  0.46 \\
V-MA          &   1.0 & 0.53  &  0.33 \\
GE-BA         &   1.0 & 0.20  &  0.34 \\
OXY-XOM       &   1.0 & 0.77  &  0.22 \\
WELL-VTR      &   1.0 & 0.99  &  0.98 \\
PPG-SHW       &   1.0 & 0.33  &  0.12 \\
VZ-TMUS       &   1.0 & 0.10  &  0.01 \\
CSX-NSC       &   1.0 & 0.12  &  0.14 \\
\bottomrule
\end{tabular} 
\caption{Comparison of pairs coefficients derived from various methods.}
\label{tab2} 
\end{table}

\begin{figure}[H]
\subfigure[MSFT-GOOGL]{\includegraphics[width=3.8 cm]{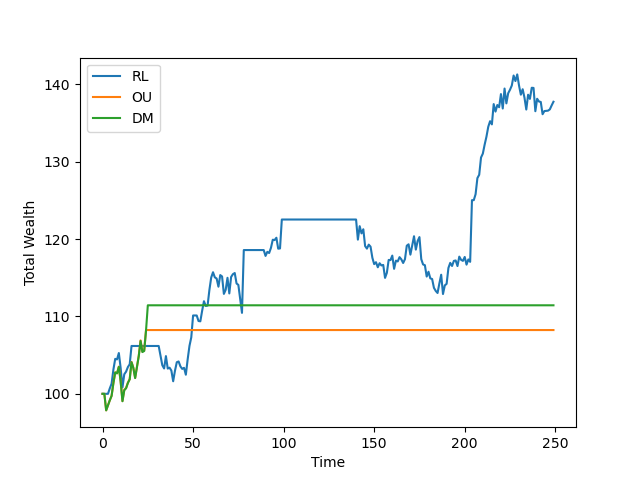}}
\subfigure[CVS-JNJ]{\includegraphics[width=3.8 cm]{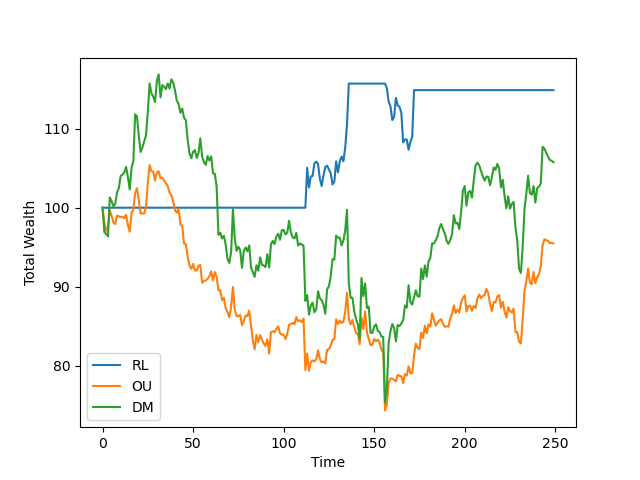}}
\subfigure[CL-KMB]{\includegraphics[width=3.8 cm]{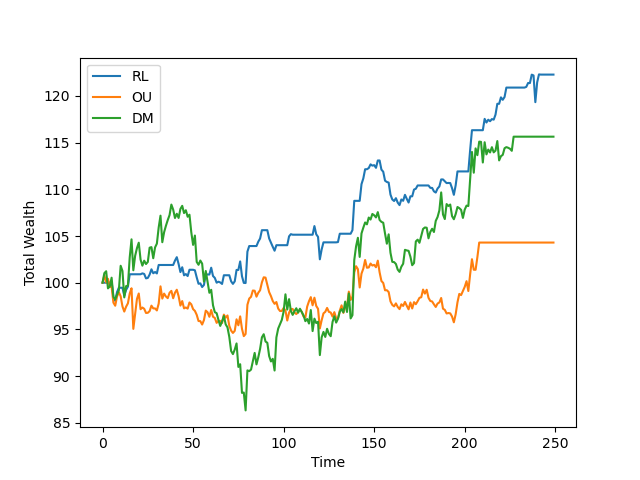}}
\subfigure[V-MA]{\includegraphics[width=3.8 cm]{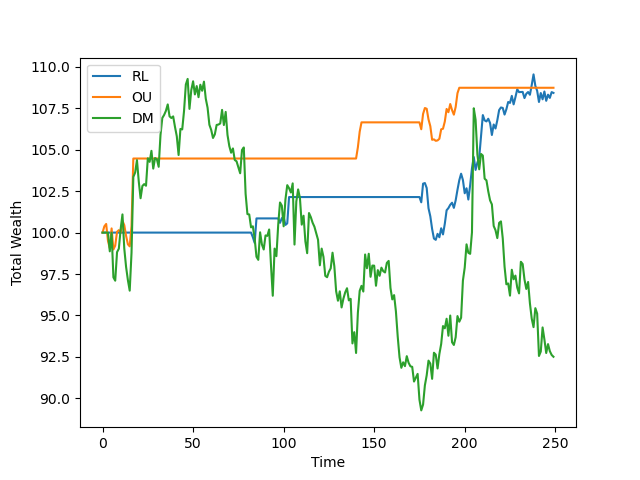}}\\
\subfigure[GE-BA]{\includegraphics[width=3.8 cm]{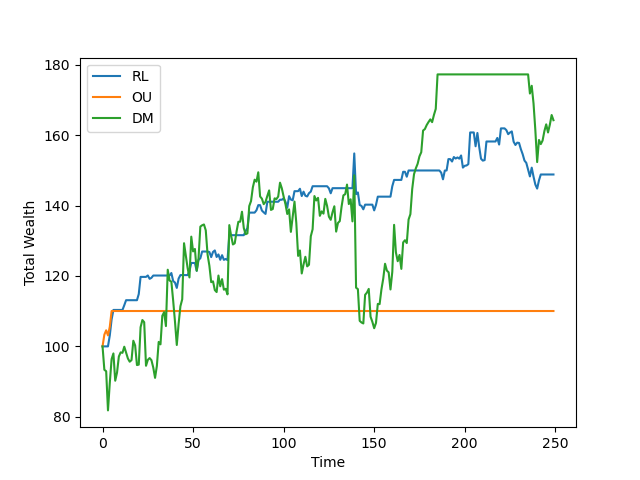}}
\subfigure[OXY-XOM]{\includegraphics[width=3.8 cm]{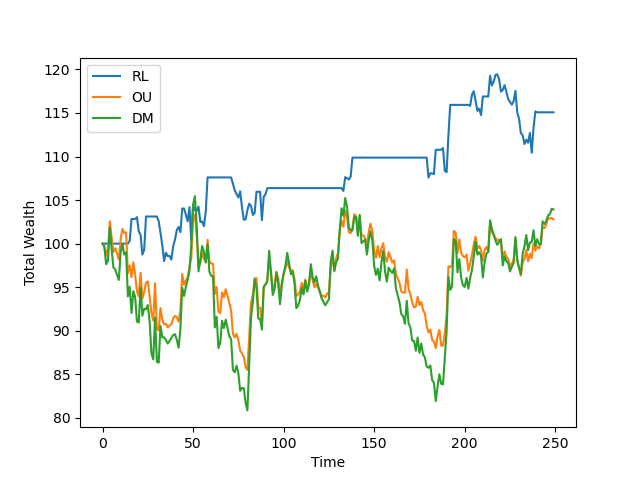}}
\subfigure[WELL-VTR]{\includegraphics[width=3.8 cm]{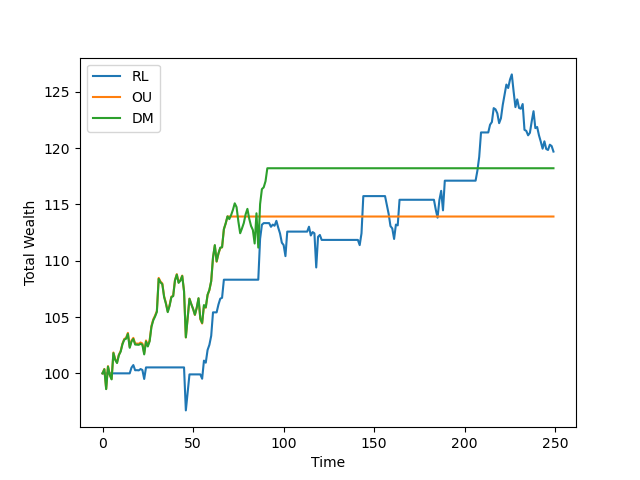}}
\subfigure[PPG-SHW]{\includegraphics[width=3.8 cm]{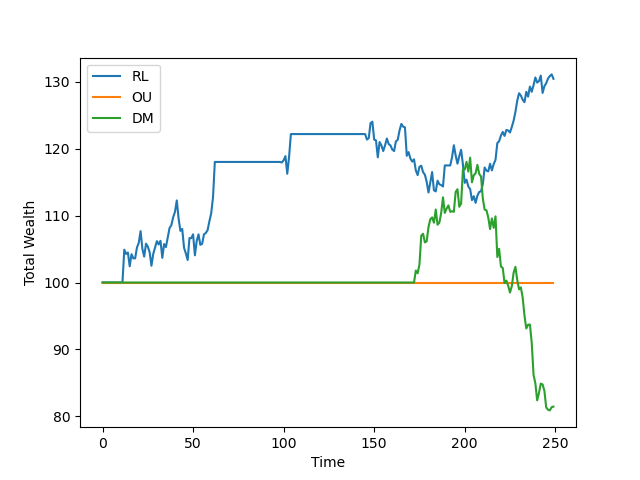}}\\
\subfigure[VZ-TMUS]{\includegraphics[width=3.8 cm]{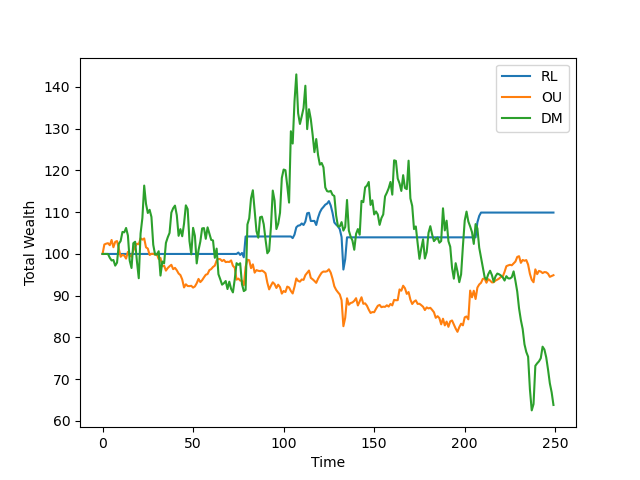}}
\subfigure[CSX-NSC]{\includegraphics[width=3.8 cm]{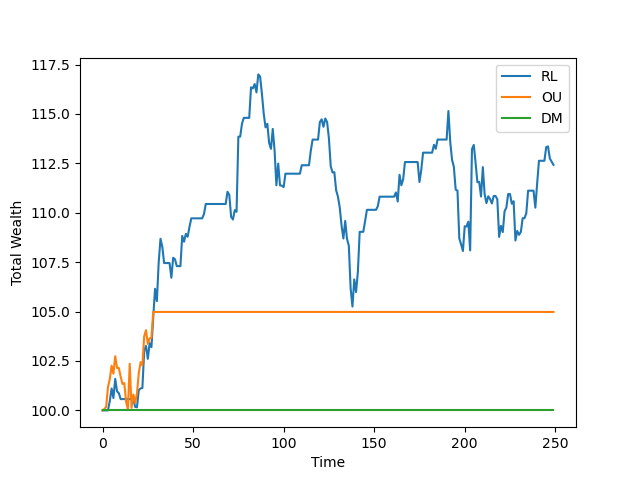}}\\
\caption{2023 total wealth growth for trading mean-reverting portfolios based on benchmarks and the proposed RL method. The initial investment is \$100.}
\label{fig3}
\end{figure}

Figure~\ref{fig3} presents the evolution of total wealth throughout the year 2023, offering an initial comparison of trading performance between the proposed method and established baselines. This visualization provides a preliminary insight into the efficacy of our strategy relative to conventional benchmarks.

The trading performance of our proposed method in comparison to established benchmarks is detailed in Tables~\ref{tab3} and~\ref{tab4}. We present a comprehensive set of performance metrics including daily returns (DailyRet), daily standard deviation (DailyStd), daily Sharpe Ratio (DailySR), maximum drawdown (MaxDD), and the annual cumulative profit and loss (CumulPnL).

Our experimental results demonstrate that the proposed reinforcement learning approach significantly outperforms traditional benchmarks in terms of daily Sharpe Ratio and cumulative returns, thereby evidencing its effectiveness and robustness in executing mean reversion trading strategies across diverse market sectors. This distinct out-performance underscores the potential benefits of incorporating reinforcement learning techniques into mean reversion trading frameworks. A pivotal factor in achieving such success is the careful design of the reinforcement learning framework, tailored to align with the specific nuances and challenges of the financial applications.

\begin{table}
\centering
\begin{tabular}{c|cccccc}
\toprule
\textbf{Index} & \textbf{MSFT-GOOGL} & \textbf{CVS-JNJ} & \textbf{CL-KMB} & \textbf{V-MA} & \textbf{GE-BA}  \\ 
\midrule
&&&  \textbf{DM Method} \\
\midrule
DailyRet ($\%$) & 0.0446  & 0.0440   & 0.0659   &  -0.0244 & 0.2771\\ 
DailyStd ($\%$) & 0.4670  & 2.0692   & 1.2314   &  1.1796  & 3.9356\\ 
DailySR 	    & 0.0955  & 0.0213   & 0.0535   &  -0.0207 & 0.0704\\
MaxDD ($\%$)    & -2.1344 & -24.6778 & -13.6791 & -10.7238 & -18.1823\\ 
CumulPnL ($\%$) & 11.4443 & 5.7581   & 15.6385  &  -7.4888 & 64.2387\\  
\midrule
&&&  \textbf{OU Method}\\
\midrule
DailyRet ($\%$) & 0.0327 & -0.0073 & 0.0198 & 0.0342 & 0.0392\\ 
DailyStd ($\%$) & 0.4285 &  1.4950 & 0.7589 & 0.3475 & 0.3890 \\ 
DailySR 	   & 0.0764  & -0.0049 & 0.0261 & 0.0985 & 0.1007\\
MaxDD ($\%$)   &-2.1427  & -25.6665 & -5.7253 & -1.0185 & 0.000 \\ 
CumulPnL ($\%$)   & 8.2443 & -4.5179 & 4.298 & 8.7348   & 10.046\\  
\midrule
&&&  \textbf{RL Method}\\
\midrule
DailyRet ($\%$) & 0.1344  & 0.0585  &  0.0826 & 0.0330 & 0.1679 \\ 
DailyStd ($\%$) & 1.0754  & 0.7506  & 0.6000 & 0.3144 & 1.2803 \\ 
DailySR 	    & 0.1250  & 0.0780  & 0.1377 &  0.1049 & 0.1312 \\
MaxDD ($\%$)    & 0.0000  & 0.0000  & -1.9476 & -0.6211 &  0.0000 \\ 
CumulPnL ($\%$) & 37.7555 & 14.8895 & 22.2879 & 8.4248 & 48.8196 \\ 
\bottomrule
\end{tabular} 
\caption{Performance summary for trading mean reversion portfolios by baselines and the proposed RL method.}
\label{tab3} 
\end{table}

\begin{table}
\centering
\begin{tabular}{c|ccccc}
\toprule
\textbf{Index} & \textbf{OXY-XOM} & \textbf{WELL-VTR} & \textbf{PPG-SHW} & \textbf{VZ-TMUS} & \textbf{CSX-NSC}  \\ 
\midrule
&&& \textbf{DM Method}\\
\midrule
DailyRet ($\%$) &  0.0373 & 0.0694 & -0.0772  & -0.112     & 0.0000 \\ 
DailyStd ($\%$) &  2.0950 & 0.6555 & 1.0113   & 3.7311     & 0.0000 \\ 
DailySR 	   &   0.0178 & 0.1058 & -0.0764  & -0.0300    & 0.0000 \\
MaxDD ($\%$)   & -19.1535 & -1.4004 & -19.1196 & -37.4779  & 0.0000 \\ 
CumulPnL ($\%$)  & 3.9194 & 18.2114 & -18.5547 & -36.1756  & 0.0000 \\  
\midrule
&&& \textbf{OU Method} \\
\midrule
DailyRet ($\%$) & 0.0238   & 0.0539  & 0.0000 & -0.0123 & 0.0199 \\ 
DailyStd ($\%$) & 1.6012   & 0.5480  & 0.0000 & 1.3241  & 0.2879 \\ 
DailySR 	    & 0.0149   & 0.0983  & 0.0000 & -0.0093 & 0.0693 \\
MaxDD ($\%$)    & -14.5001 & -1.3798 & 0.0000 & -18.652 & 0.0000 \\ 
CumulPnL ($\%$) & 2.7812   & 13.9245 & 0.0000 & -5.0869 & 4.9825 \\  
\midrule
&&& \textbf{RL Method}\\
\midrule
DailyRet ($\%$) & 0.0609  & 0.0745  & 0.1124   &  0.0412  & 0.0496 \\ 
DailyStd ($\%$) & 0.9446  & 0.6794  & 1.0600   &  0.8037  & 0.7101 \\ 
DailySR 	    & 0.0861  & 0.1097  & 0.1061   &  0.0513  & 0.0698 \\
MaxDD ($\%$)    & -2.0008 & -3.2895 & 0.0000   &  -3.7306 & 0.0000 \\ 
CumulPnL ($\%$) & 15.0791 & 19.6910 & 30.4559  &  9.9163  & 12.4263 \\ 
\bottomrule
\end{tabular} 
\caption{Performance summary for trading mean reversion portfolios by baselines and the proposed RL method.}
\label{tab4} 
\end{table}

\section{Conclusions and Future plan}
This study has presented a novel approach to statistical arbitrage by integrating a model-free framework with reinforcement learning techniques. By establishing an empirical mean reversion time metric and optimizing asset coefficients to minimize this duration, our work has substantially refined the process of constructing mean reversion spreads. Furthermore, we have formulated a reinforcement learning framework for the trading phase, carefully designing the state space to encapsulate the recent trends in price movements and the reward functions to align with the distinct attributes of mean reversion trading. The empirical analysis conducted over the several sectors within the US market has underscored the proposed method's effectiveness and consistency.

For future work, we aim to explore more sophisticated reinforcement learning algorithms to further optimize trading strategies. This will include the application of deep reinforcement learning and exploration of various reward structures to enhance strategy performance.

\bibliographystyle{plainnat}
\bibliography{main}

\begin{thebibliography}{21}
\providecommand{\natexlab}[1]{#1}
\providecommand{\url}[1]{\texttt{#1}}
\expandafter\ifx\csname urlstyle\endcsname\relax
  \providecommand{\doi}[1]{doi: #1}\else
  \providecommand{\doi}{doi: \begingroup \urlstyle{rm}\Url}\fi

\bibitem[Avellaneda and Lee(2010)]{avellaneda2010statistical}
Marco Avellaneda and Jeong-Hyun Lee.
\newblock Statistical arbitrage in the us equities market.
\newblock \emph{Quantitative Finance}, 10\penalty0 (7):\penalty0 761--782, 2010.

\bibitem[Chang et~al.(2021)Chang, Man, Xu, and Hsu]{chang2021pairs}
Victor Chang, Xiaowen Man, Qianwen Xu, and Ching-Hsien Hsu.
\newblock Pairs trading on different portfolios based on machine learning.
\newblock \emph{Expert Systems}, 38\penalty0 (3):\penalty0 e12649, 2021.

\bibitem[Do and Faff(2012)]{do2012pairs}
Binh Do and Robert Faff.
\newblock Are pairs trading profits robust to trading costs?
\newblock \emph{Journal of Financial Research}, 35\penalty0 (2):\penalty0 261--287, 2012.

\bibitem[Do et~al.(2006)Do, Faff, and Hamza]{do2006new}
Binh Do, Robert Faff, and Kais Hamza.
\newblock A new approach to modeling and estimation for pairs trading.
\newblock In \emph{Proceedings of 2006 financial management association European conference}, volume~1, pages 87--99. Citeseer, 2006.

\bibitem[Elliott et~al.(2005)Elliott, Van Der~Hoek, and Malcolm]{elliott2005pairs}
R.J. Elliott, J.~Van Der~Hoek, and W.P. Malcolm.
\newblock Pairs trading.
\newblock \emph{Quantitative Finance}, 5\penalty0 (3):\penalty0 271--276, 2005.

\bibitem[Engle and Granger(1987)]{engle1987co}
Robert~F Engle and Clive~WJ Granger.
\newblock Co-integration and error correction: representation, estimation, and testing.
\newblock \emph{Econometrica}, 55\penalty0 (2):\penalty0 251--276, 1987.

\bibitem[Fink and Gandhi(2007)]{fink2007important}
Eugene Fink and Harith~Suman Gandhi.
\newblock Important extrema of time series.
\newblock 2007.

\bibitem[Galenko et~al.(2012)Galenko, Popova, and Popova]{galenko2012trading}
Alexander Galenko, Elmira Popova, and Ivilina Popova.
\newblock Trading in the presence of cointegration.
\newblock \emph{The Journal of Alternative Investments}, 15\penalty0 (1):\penalty0 85--97, 2012.

\bibitem[Gatev et~al.(2006)Gatev, Goetzmann, and Rouwenhorst]{gatev2006pairs}
Evan Gatev, William~N Goetzmann, and K~Geert Rouwenhorst.
\newblock Pairs trading: Performance of a relative-value arbitrage rule.
\newblock \emph{Review of Financial Studies}, 19\penalty0 (3):\penalty0 797--827, 2006.

\bibitem[Guijarro-Ordonez et~al.(2021)Guijarro-Ordonez, Pelger, and Zanotti]{guijarro2021deep}
Jorge Guijarro-Ordonez, Markus Pelger, and Greg Zanotti.
\newblock Deep learning statistical arbitrage.
\newblock \emph{Available at SSRN 3862004}, 2021.

\bibitem[Huck and Afawubo(2015)]{huck2015pairs}
Nicolas Huck and Komivi Afawubo.
\newblock Pairs trading and selection methods: is cointegration superior?
\newblock \emph{Applied Economics}, 47\penalty0 (6):\penalty0 599--613, 2015.

\bibitem[Kanamura et~al.(2010)Kanamura, Rachev, and Fabozzi]{kanamura2010profit}
T.~Kanamura, S.T. Rachev, and F.J. Fabozzi.
\newblock A profit model for spread trading with an application to energy futures.
\newblock \emph{The Journal of Trading}, 5\penalty0 (1):\penalty0 48--62, 2010.

\bibitem[Leung and Li(2015)]{leung2015optimal}
Tim Leung and Xin Li.
\newblock Optimal mean reversion trading with transaction costs and stop-loss exit.
\newblock \emph{International Journal of Theoretical and Applied Finance}, 18\penalty0 (03):\penalty0 1550020, 2015.

\bibitem[Leung and Li(2016)]{meanReversionBook}
Tim Leung and Xin Li.
\newblock \emph{Optimal Mean Reversion Trading: Mathematical Analysis and Practical Applications}.
\newblock Modern Trends in Financial Engineering. World Scientific Publishing Company, 2016.

\bibitem[Leung and Nguyen(2019)]{leung2019constructing}
Tim Leung and Hung Nguyen.
\newblock Constructing cointegrated cryptocurrency portfolios for statistical arbitrage.
\newblock \emph{Studies in Economics and Finance}, 2019.

\bibitem[Liew and Wu(2013)]{liew2013pairs}
Rong~Qi Liew and Yuan Wu.
\newblock Pairs trading: A copula approach.
\newblock \emph{Journal of Derivatives \& Hedge Funds}, 19\penalty0 (1):\penalty0 12--30, 2013.

\bibitem[Montana and Triantafyllopoulos(2011)]{Montana2011}
G.~Montana and K.~Triantafyllopoulos.
\newblock Dynamic modeling of mean reverting spreads for statistical arbitrage.
\newblock \emph{Computational Management Science}, 8:\penalty0 23--49, 2011.

\bibitem[Sarmento and Horta(2020)]{sarmento2020enhancing}
Sim{\~a}o~Moraes Sarmento and Nuno Horta.
\newblock Enhancing a pairs trading strategy with the application of machine learning.
\newblock \emph{Expert Systems with Applications}, 158:\penalty0 113490, 2020.

\bibitem[Vidyamurthy(2004)]{vidyamurthy2004pairs}
Ganapathy Vidyamurthy.
\newblock \emph{Pairs Trading: quantitative methods and analysis}, volume 217.
\newblock John Wiley \& Sons, 2004.

\bibitem[Xie et~al.(2016)Xie, Liew, Wu, and Zou]{xie2016pairs}
Wenjun Xie, Rong~Qi Liew, Yuan Wu, and Xi~Zou.
\newblock Pairs trading with copulas.
\newblock \emph{The Journal of Trading}, 11\penalty0 (3):\penalty0 41--52, 2016.

\bibitem[Zhang et~al.(2020)Zhang, Leung, and Aravkin]{zhang2020sparse}
Jize Zhang, Tim Leung, and Aleksandr Aravkin.
\newblock Sparse mean-reverting portfolios via penalized likelihood optimization.
\newblock \emph{Automatica}, 111:\penalty0 108651, 2020.

\end{thebibliography}

\end{document}